# Tunable Ultraviolet Vertically-emitting Organic Laser

Sébastien Forget[1], Hadi Rabbani-Haghighi, Nordine Diffalah, Alain Siove and Sébastien Chénais[1]

Laboratoire de Physique des Lasers, Université Paris 13 / CNRS, 99 avenue Jean-Baptiste Clément, 93430 Villetaneuse, France

**Abstract :**

**A solid-state organic thin-film laser with intracavity frequency doubling is reported. Tunable ultraviolet emission from 309 to 322 nm is achieved from a vertical external cavity surface-emitting organic laser, with 2 % efficiency (1 µJ at 315 nm). The laser comprises a polymethyl(methacrylate) layer doped with Rhodamine 640, spun-cast onto a plane mirror, a remote concave mirror, a nonlinear crystal and a dichroic separator. The output is spectrally narrow (<0.5 nm FWHM) and tunable through phase-matching selection of the fundamental radiation lasing modes. These results highlight a low-cost and portable alternative to tunable UV laser sources, useful for spectroscopic applications.**

---

[1] Author to whom correspondence should be addressed

Tunable UV (200-400 nm) laser sources are of fundamental interest in many fields such as atmospheric spectroscopy, ionization spectrometry, chemical or biological hazard detection, laser-induced fluorescence spectroscopy, combustion diagnostics, or photobiology [1].

For most of these applications, portable and low-cost devices are desirable. Tunable UV radiation has been demonstrated using several routes: for instance upon using nonlinear optical conversion, either by optical parametric oscillation [2-4] or sum-frequency mixing from dye lasers [5], Ti-sapphire lasers, alexandrite [6] or Cr-doped colquiriites [7]. It may also be obtained directly from rare-earth-doped crystals pumped by the fourth or fifth harmonic of a Nd laser [8]. All these systems remain expensive and hardly transportable.

In the other hand, a combination of broad tunability, low cost and compactness can be achieved with organic solid-state lasers [9]. This is especially true when the gain medium takes the form of a thin film, since the deposition of the gain medium onto the substrate (by *e.g.* spin casting or ink-jet printing) can be a high-throughput and cost-effective process, balancing the drawbacks of a finite lifetime. Furthermore, many demonstrations of lasing in organic semiconductors have presented the possibility for realizing electrically-pumped lasers based on these materials [10]; but while this goal has not yet been reached, recent reports of diode-pumping [11] and LED-pumping [12] suggested that cheap tunable optically-pumped lasers may soon become a reality.

However, the tunability of organic pi-conjugated systems is limited around the visible spectrum, and extending the remarkable properties of organic lasers to lower wavelengths is fundamentally restricted to the near-UV domain. UV dyes have indeed little-extended pi-conjugated systems and are inefficient light emitters in virtue of their low quantum yield of fluorescence [13]; furthermore the deep-UV optical excitation needed to obtain lasing in the UV results in a very fast photodegradation. Therefore, among a few reports of organic lasers below 400 nm [14,15], the lowest wavelength obtained to date directly from an organic thin-film laser is 361 nm [16].

We propose an alternative route to reach the UV domain with an organic thin-film device, allowing lower wavelengths to be attained (potentially down to the UV-C range), while not compromising photostability: it consists in using intracavity frequency conversion in an open-cavity thin-film organic laser resonator. Whereas frequency doubling has been commonly used in liquid dye lasers to obtain tunable UV light [17], it has remained largely unexplored in solid-state dye lasers, presumably because available intensities were too weak. External frequency doubling has been reported from a bulk dye-doped rotating plastic disk laser [18], but has never been demonstrated in a thin-film organic laser.

We have based the design of an efficient frequency-doubled organic laser on two simple building principles: 1) the laser beam at the fundamental wavelength must have high peak brightness, *i.e.* high peak power and high beam quality altogether; and 2) the resonator must be open to enable intracavity mixing, upon insertion of an efficient nonlinear crystal inside the cavity. Neither condition is actually met in traditional organic thin-film laser structures, such as DFB or DBR lasers, microcavities and microdisks [16,19-21]. It has been proposed to mix an organic nonlinear chromophore with an active lasing dye within the same structure, however the concept requires complex poling techniques and has been up to now only demonstrated with a liquid dye [22].

In this letter, we propose a hybrid structure composed of a red-emitting thin-film open-resonator laser comprising an intracavity beta barium borate (BBO) crystal phase-matched for frequency doubling and a harmonic separator. This architecture derives from the VECSOL concept (Vertical external Cavity Surface-emitting Organic Laser) recently demonstrated in our group [23]. The VECSOL is the organic counterpart of the Vertical External Cavity Surface Emitting Laser[24,25] and is made up of a thin layer of a dye-doped polymer spun-cast onto a high-reflectivity mirror, and a remote concave output coupler closing the cavity. The structure is optically pumped by a frequency doubled Nd: YAG laser operating at 10 Hz with 7-ns long pulses. The pump spot diameter on the organic sample was set to 180 µm in order to match the fundamental cavity mode size. High conversion efficiencies were obtained (up to 57 %, corresponding to output peak powers in the kW range, and intracavity peak powers of several tens of kW), together with excellent beam quality ($M^2=1$, see ref. [23] for details). The open cavity of VECSEL makes it the ideal structure to explore intracavity frequency mixing [25].

The experimental setup is described and illustrated in figs. 1 and 2. The BBO crystal (section 10x10 mm) for frequency doubling was cut for type-I phase matching at 620 nm (Θ=38.9°, ϕ= 90°). The thickness of the crystal was 3 mm, determined as a trade-off between nonlinear conversion efficiency and the walk-off effect, which is quite large in highly birefringent BBO. The laser medium is a 21.7-µm-thick film made of 1-wt. % of Rhodamine 640 dispersed in polymethyl methacrylate (PMMA, molar weight $9.5\ 10^5$) directly spun cast onto the highly-reflective (R=99.5% @600-650 nm) plane mirror. All experiments (including spin coating) were performed in air at room temperature. The dye-doped sample absorbed 85% of the incident pump power at 532 nm. In order to maximize the intracavity peak power for the fundamental radiation, both mirrors were chosen highly reflective in the 600-650 nm range, and the concave mirror was also coated for high reflectivity in the 300-330 nm range to enable conversion of the red beam in both propagating directions. The VECSOL efficiency drops as the cavity length increases, since the oscillation buildup time increases accordingly and becomes at some point too high to enable steady state to be reached within the pump pulse [23]. Here the total length of the cavity was set to 10 mm, which was the minimum length practically achievable given the size of intracavity elements. The setup remains consequently extremely compact (see photo in the inset of fig. 2.) and could be further miniaturized.

A dichroic plate was inserted at Brewster angle inside the 10-mm-long cavity, with one face having a high-reflectivity coating in the UV (in the range 300-330 nm) for s-polarization along with a high p-polarization transmission coating in the 600-650 nm range. The role of this plate was to allow an efficient type-I phase matching scenario by forcing the red laser emission to be p-polarized, and also enabled an efficient outcoupling of the UV beam while making sure that UV photons never met the organic film. This avoids any UV-enhanced photodegradation, and decouples degradation issues from UV generation.

The input-output characteristics of the source are shown in figure 3; the red and UV emissions exhibit a linear and parabolic evolution vs. pump power, respectively, as expected when pump depletion is negligible. The UV output is 1 µJ per pulse, corresponding to a peak power of 250 W; the optical-to-optical conversion efficiency (UV energy divided by the absorbed pump energy) is 2%. This is similar to the maximum conversion efficiency of 2% obtained in an inorganic intracavity frequency-doubled VECSEL with GaInP quantum wells running in the CW mode[25]. For higher pump power levels, photodegradation provoked a sub-linear behavior of the laser efficiency curve (see the last points in fig. 3) and a reduced lifetime. The spatial profile of the UV beam is shown in the inset of the figure 3. The fundamental laser beam is diffraction limited, and the slight anisotropy of the UV spot is due to the small angular acceptance of BBO.

The spectrum of the red laser is controlled by the sub-cavity formed by the organic thin film acting as a Fabry-Perot etalon (see figure 4); it comprises several peaks spaced by 6 nm corresponding to the free spectral range of a 21.7-µm thick etalon. Each peak contains many modes of the external cavity, which cannot be resolved with a standard spectrometer. The UV spectrum, in contrast, exhibits a single peak (with a spectral width below 0.5 nm - limited by the SPEX 270M spectrometer resolution), because the spectral acceptance of the nonlinear process (also on the order of 0.5 nm) allows only one of the red peaks to fulfill the phase-matching condition at a given crystal angle. Upon tilting the BBO crystal by a fraction of degree, it is straightforward to make the spectral acceptance curve matching any of the red peaks. Moreover, we observed non-degenerate sum-frequency mixing occurring between two adjacent red peaks (for example, 626nm + 632 nm = 314.5 nm), which leads to additional UV peaks located in-between frequency-doubled lines. In fig. 4 a typical series of spectra is shown corresponding to a given position of the pump spot onto the sample; it corresponds to a UV emission tunable by discrete steps upon BBO orientation from 309.5 to 316 nm. Interestingly, the device is also continuously tunable between these values in virtue of the non-perfect thickness homogeneity of the film, especially near the edges where the spin-casting process causes the film to be thicker than at the center [23]. It was then possible to obtain continuous tunability from 309 to 322 nm, by combining a scan of the pump spot location to a fine control of BBO orientation.

At last, we studied the laser photostability at a pump energy fixed at 4 times the laser threshold (figure 5.) More than 5100 shots were observed before the UV intensity dropped to half its initial value. A full scan of a 1-inch sample can provide hundreds of hours of UV laser light with energy between 2 and 0.2 µJ per pulse. It can be seen in fig. 5 that the UV output does not vary with time as the square of the fundamental red energy, as it should be the case since we expect the dichroic plate to protect the sample from a direct UV degradation, as mentioned above. We relate this effect to the slight blueshift observed on the fundamental output (see inset of fig. 5) induced by refractive index modifications in the sample. A shift from 650 to 649.2 nm was measured within one hour of operation (36000 pulses) when the pump level was 4 times above threshold. As the dye molecules degrade, the index of refraction decreases towards the lower index of undoped PMMA, causing the resonance condition of the Fabry-Perot etalon to shift, and the lasing wavelength accordingly, inducing a phase mismatch. Taking into account the wavelength shift in the spectral acceptance of BBO, we predicted a UV emission behavior from the red emission that fits correctly the experimental data (see fig. 5). A slight realignment of the nonlinear crystal brings the UV energy back to a higher value which confirms that the accelerated degradation rate is only due to phase matching issues. An active tracking system could then easily be implemented to restore the expected degradation profile. The UV emission lifetime can therefore be considered as only limited by the lifetime of the red dye, which can be extremely high (millions of pulses) in red fluorescent dyes. The extended lifetime is at the cost of a deriving wavelength, which is not necessarily a problem in applications such as biology where spectral signatures are broad.

In conclusion, we demonstrated an ultraviolet organic thin-film laser tunable between 309 and 322 nm, based on intracavity sum-frequency mixing of Rhodamine 640 in PMMA by an inorganic BBO frequency converter. The device is based on the Vertical External cavity Surface-emitting Organic Laser (VECSOL) concept, which produces high-brightness beams and offers an open cavity allowing for the implementation of intracavity nonlinear conversion. UV energies in the µJ range were obtained with a 2% optical-to-optical conversion efficiency. The spectrum of the UV output contains a single peak (<0.5 nm FWHM), since spectral acceptance of BBO plays the role of a narrowband filter on the multiple-peak spectrum of the fundamental red laser. Wavelength tuning was obtained through phase-matching control (upon BBO orientation, by discrete steps) as well as through layer thickness control (uncontrolled fluctuations near the edges), resulting in a continuous tuning range of 13 nm.

The authors acknowledge the ANR (Young researchers program) for funding this work.

# FIGURES

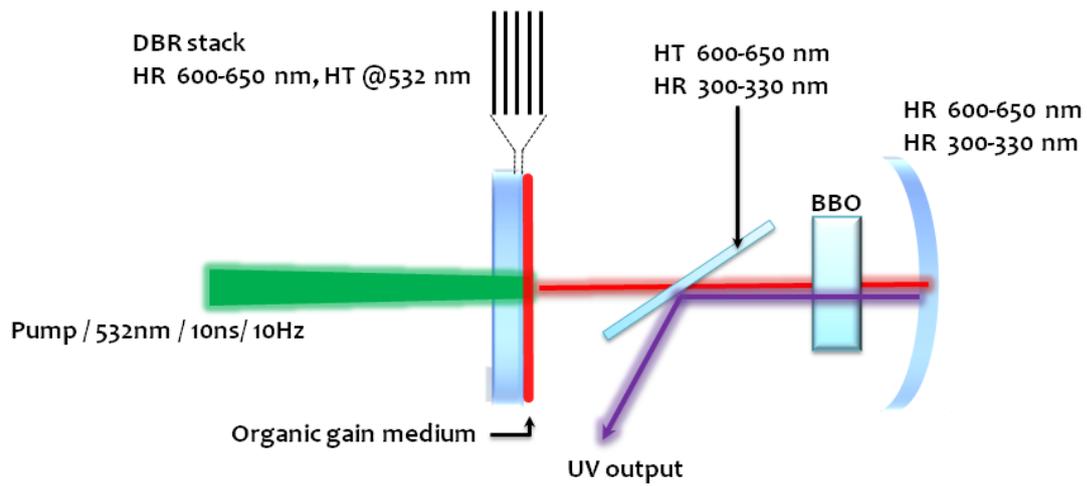

**Fig. 1**: experimental setup. The organic layer is spun cast onto the plane mirror; the concave mirror is highly reflective for both red and UV beams and has a 150-mm radius of curvature.

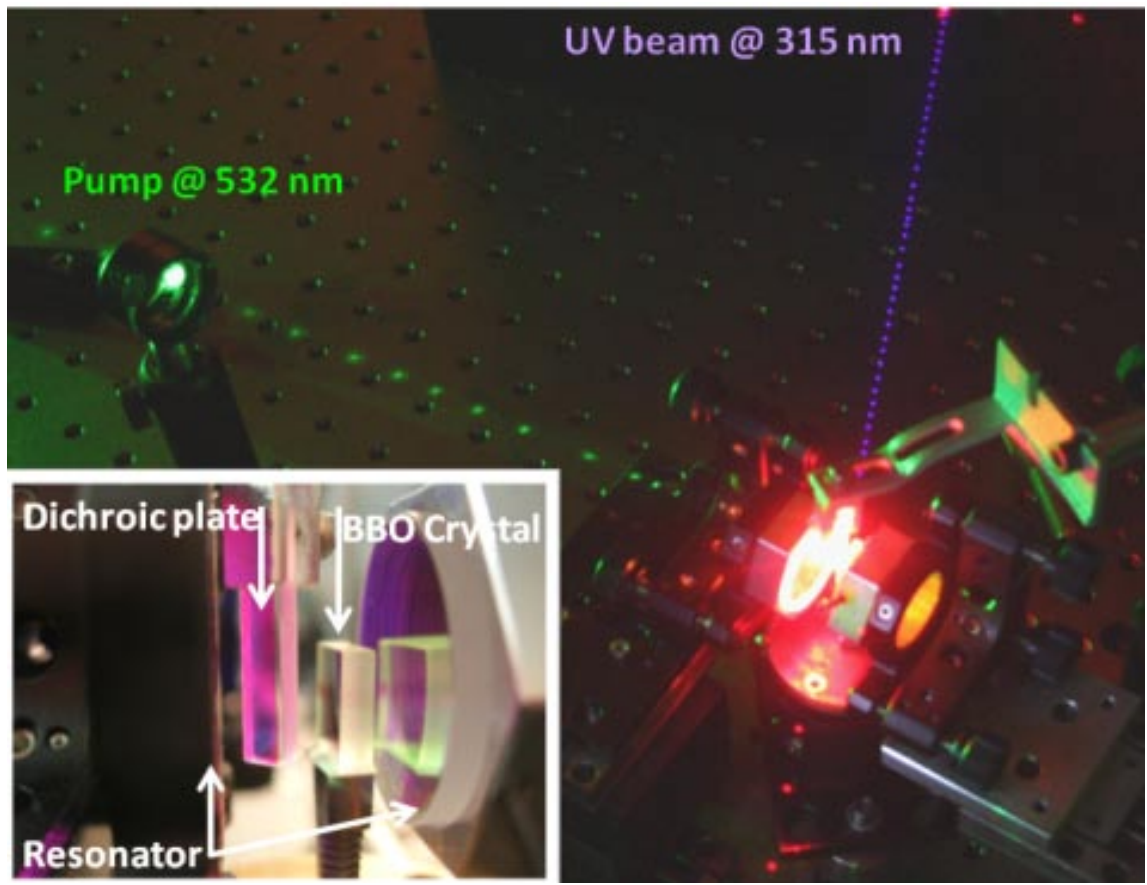

**Fig. 2** : long-exposure photograph of the setup showing the pump beam (coming from the left), a faint red beam leaking from the dichroic plate (bottom) and the UV beam, visible through fluorescence of a blank paper sheet (on top). Insert : close up of the cavity with dichroic plate and BBO crystal. The length of the cavity is 10 mm.

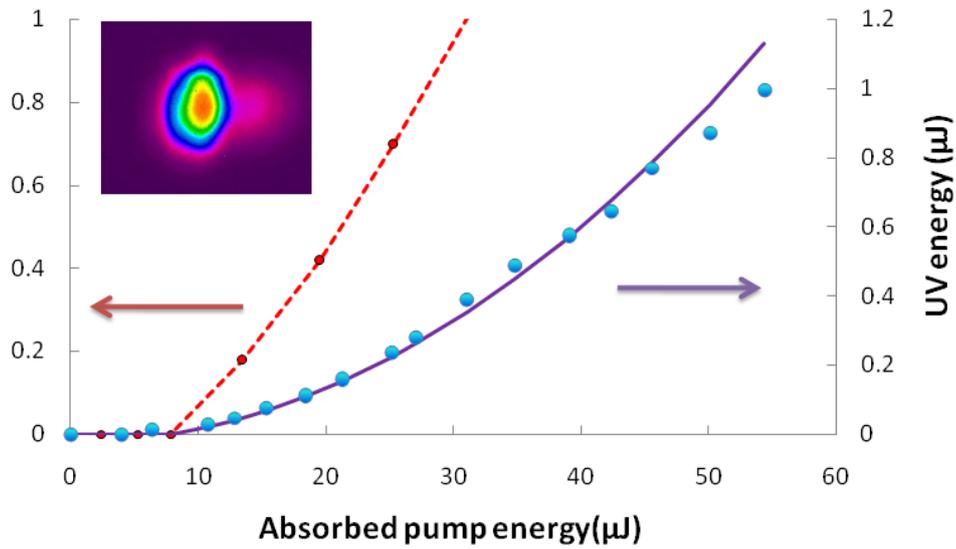

**Fig. 3**. *Left axis*: Red (fundamental) laser emission as a function of the pump energy. The structure is not designed to produce red radiation, thus the vertical axis represents one of the leaks in arbitrary units. *Right axis*: Energy of the UV output (at 315 nm) versus absorbed laser pump energy at 532 nm. The full line is a parabolic fit started from laser threshold. *Inset*: UV beam profile. The small acceptance angle of BBO causes the beam to be slightly elliptical.

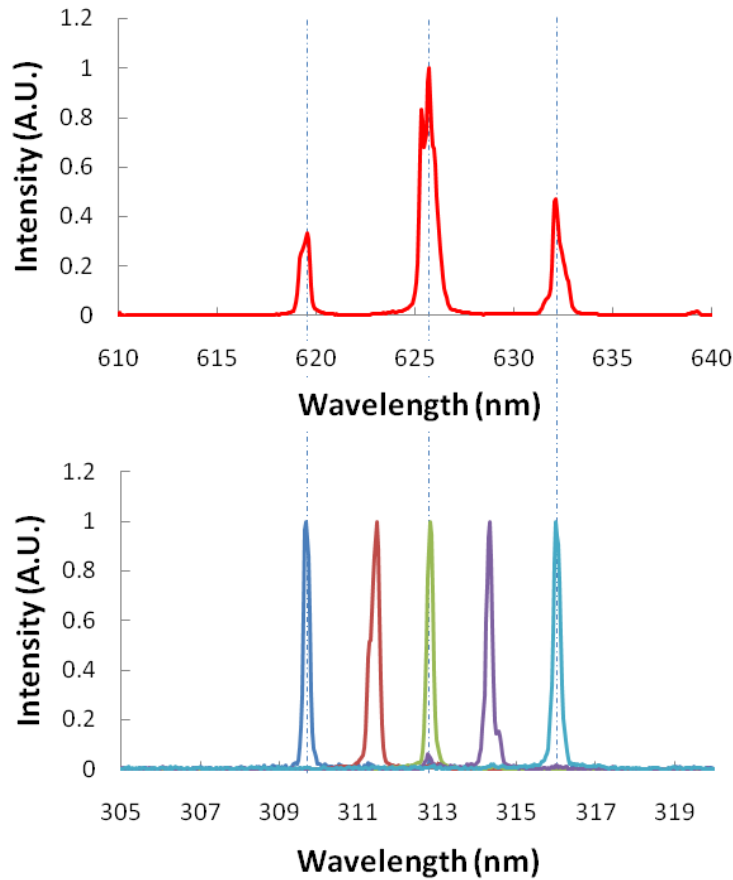

**Fig.4** : *Top:* typical emission spectrum of the laser. The spectral gap petween each peak is around 6 nm, corresponding to the free spectral range of a 21.7-µm thick organic layer. *Bottom*: several individual UV peaks obtained for a given pump spot location onto the organic layer, for different orientations of the BBO crystal (arbitrary units, normalized). In addition to second harmonic generation associated to the red peaks, non-degenerate sum-frequency mixing of adjacent red peaks occurs and produces extra UV wavelengths between the SHG peaks.

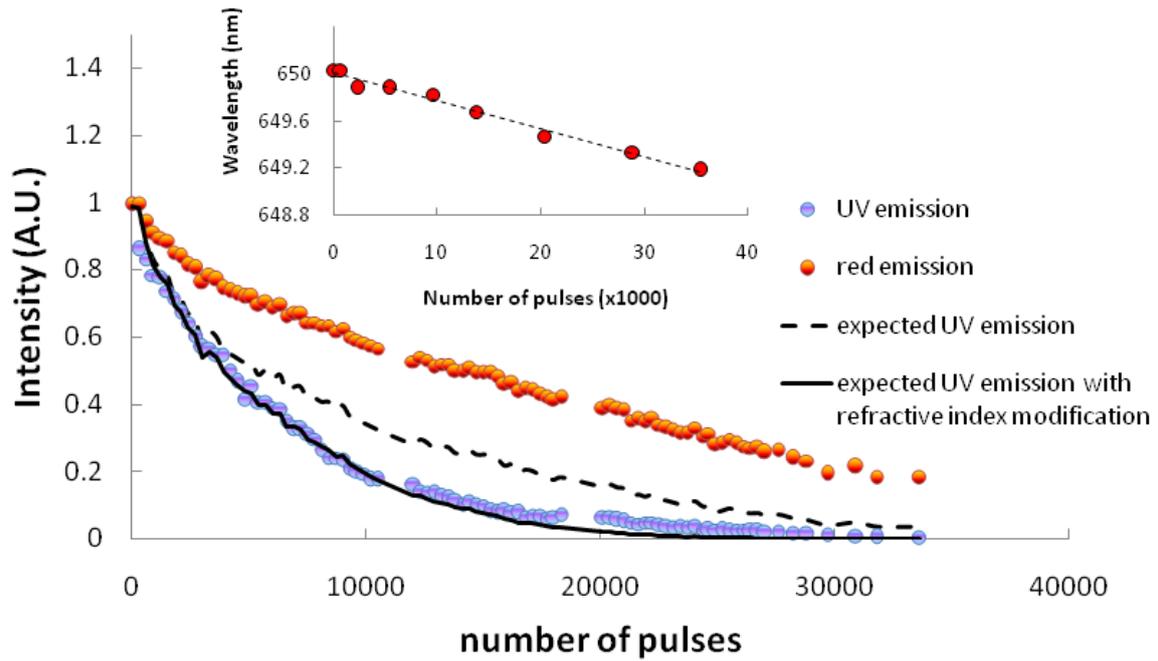

**Fig. 5**: photodegradation curves measured 4 times above laser threshold. The dotted black line represents the expected UV intensity (square of measured red intensity). The full black line is the expected UV intensity calculated from the measured red curve with the account of phase matching alteration due to the modification of the laser medium refractive index. The latter is deduced from the shift of the laser wavelength vs. time (*Inset*).